\documentclass[sn-mathphys,Numbered]{sn-jnl}

\usepackage{graphicx}%
\usepackage{multirow}%
\usepackage{amsmath,amssymb,amsfonts}%
\usepackage{amsthm}%
\usepackage{mathrsfs}%
\usepackage[title]{appendix}%
\usepackage{xcolor}%
\usepackage{textcomp}%
\usepackage{manyfoot}%
\usepackage{booktabs}%
\usepackage{algorithm}%
\usepackage{algorithmicx}%
\usepackage{algpseudocode}%
\usepackage{listings}%
\usepackage{upgreek}%
\usepackage{soul}
\soulregister\cite7
\soulregister\ref7

\theoremstyle{thmstyleone}%
\theoremstyle{thmstyletwo}%

\theoremstyle{thmstylethree}%

\begin{document}

\title[Article Title]{Simulation Studies for the First Pathfinder of the CATCH Space Mission}


\author[1,2]{\fnm{Yiming} \sur{Huang}}
\author[1]{\fnm{Juan} \sur{Zhang}}\email{zhangjuan@ihep.ac.cn}
\author[1]{\fnm{Lian} \sur{Tao}}\email{taolian@ihep.ac.cn}
\author[1]{\fnm{Zhengwei} \sur{Li}}
\author[3]{\fnm{Donghua} \sur{Zhao}}
\author[1]{\fnm{Qian-Qing} \sur{Yin}}
\author[1]{\fnm{Xiangyang} \sur{Wen}}
\author[1,2]{\fnm{Jingyu} \sur{Xiao}}
\author[3]{\fnm{Chen} \sur{Zhang}}
\author[1,2]{\fnm{Shuang-Nan} \sur{Zhang}}
\author[1]{\fnm{Shaolin} \sur{Xiong}}

\author[4]{\fnm{Qingcui} \sur{Bu}}
\author[5]{\fnm{Jirong} \sur{Cang}}
\author[6]{\fnm{Dezhi} \sur{Cao}}
\author[7]{\fnm{Wen} \sur{Chen}}
\author[7]{\fnm{Siran} \sur{Ding}}
\author[1]{\fnm{Min} \sur{Gao}}
\author[7]{\fnm{Yang} \sur{Gao}}
\author[8]{\fnm{Shujin} \sur{Hou}}
\author[5]{\fnm{Liping} \sur{Jia}}
\author[9]{\fnm{Ge} \sur{Jin}}
\author[10]{\fnm{Dalin} \sur{Li}}
\author[7]{\fnm{Jinsong} \sur{Li}}
\author[1,2]{\fnm{Panping} \sur{Li}}
\author[1,2]{\fnm{Yajun} \sur{Li}}
\author[1]{\fnm{Xiaojing} \sur{Liu}}
\author[1,2]{\fnm{Ruican} \sur{Ma}}
\author[5]{\fnm{Xingyu} \sur{Pan}}
\author[1]{\fnm{Liqiang} \sur{Qi}}
\author[1,11]{\fnm{Jinhui} \sur{Rao}}
\author[7]{\fnm{Xianfei} \sur{Sun}}
\author[11]{\fnm{Qingwen} \sur{Tang}}
\author[12]{\fnm{Ruijing} \sur{Tang}}
\author[1]{\fnm{Yusa} \sur{Wang}}
\author[7]{\fnm{Yibo} \sur{Xu}}
\author[1]{\fnm{Sheng} \sur{Yang}}
\author[1]{\fnm{Yanji} \sur{Yang}}
\author[7]{\fnm{Yong} \sur{Yang}}
\author[1,11]{\fnm{Xuan} \sur{Zhang}}
\author[7]{\fnm{Yueting} \sur{Zhang}}
\author[7]{\fnm{Heng} \sur{Zhou}}
\author[1,2]{\fnm{Kang} \sur{Zhao}}
\author[1,2]{\fnm{Qingchang} \sur{Zhao}}
\author[1,2]{\fnm{Shujie} \sur{Zhao}}
\author[1]{\fnm{Zijian} \sur{Zhao}}

\affil[1]{\orgdiv{Key Laboratory for Particle Astrophysics, Institute of High Energy Physics}, \orgname{Chinese Academy of Sciences}, \orgaddress{\city{Beijing}, \postcode{100149}, \country{China}}}

\affil[2]{\orgdiv{University of Chinese Academy of Sciences}, \orgname{Chinese Academy of Sciences}, \orgaddress{\city{Beijing}, \postcode{100149}, \country{China}}}

\affil[3]{\orgdiv{National Astronomical Observatories}, \orgname{Chinese Academy of Sciences}, \orgaddress{\city{Beijing}, \postcode{100101}, \country{China}}}

\affil[4]{\orgdiv{Institut f\"ur Astronomie und Astrophysik, Kepler Center for Astro and Particle Physics}, \orgname{Eberhard Karls Universi\"at}, \orgaddress{\street{Sand 1},\city{\"ubingen}, \postcode{72076}, \country{Germany}}}

\affil[5]{\orgname{Star Detect CO.}, \orgname{LTD.}, \orgaddress{\city{Beijing}, \postcode{100190}, \country{China}}}

\affil[6]{\orgname{Tsinghua University}, \orgaddress{\city{Beijing}, \postcode{100084}, \country{China}}}

\affil[7]{\orgname{Innovation Academy for Microsatellites of Chinese Academy of Sciences}, \orgaddress{\city{Shanghai}, \postcode{200135}, \country{China}}}

\affil[8]{\orgname{Nanyang Normal University}, \orgaddress{\city{Nanyang}, \postcode{473061}, \country{China}}}

\affil[9]{\orgdiv{North Night Vision Technology CO.}, \orgname{LTD.}, \orgaddress{\city{Nanjing}, \postcode{211106}, \country{China}}}

\affil[10]{\orgdiv{National Space Science Center}, \orgname{Chinese Academy of Sciences}, \orgaddress{\city{Beijing}, \postcode{100190}, \country{China}}}

\affil[11]{\orgname{Nanchang University}, \orgaddress{\city{Nanchang}, \postcode{330031}, \country{China}}}

\affil[12]{\orgname{Beijing Jiaotong University}, \orgaddress{\city{Beijing}, \postcode{100044}, \country{China}}}

\abstract{The Chasing All Transients Constellation Hunters (CATCH) space mission is an intelligent constellation consisting of 126 micro-satellites in three types (A, B, and C), designed for X-ray observation with the objective of studying the dynamic universe. Currently, we are actively developing the first Pathfinder (CATCH-1) for the CATCH mission, specifically for type-A satellites. CATCH-1 is equipped with Micro Pore Optics (MPO) and a 4-pixel Silicon Drift Detector (SDD) array. To assess its scientific performance, including the effective area of the optical system, on-orbit background, and telescope sensitivity, we employ the Monte Carlo software Geant4 for simulation in this study. The MPO optics exhibit an effective area of 41\,cm$^2$ at the focal spot for 1\,keV X-rays, while the entire telescope system achieves an effective area of 29 cm$^2$ at 1\,keV when taking into account the SDD detector's detection efficiency. The primary contribution to the background is found to be from the Cosmic X-ray Background. Assuming a 625\,km orbit with an inclination of $29^\circ$, the total background for CATCH-1 is estimated to be $8.13\times10^{-2}$\,counts~s$^{-1}$ in the energy range of 0.5--4\,keV. Based on the background within the central detector and assuming a Crab-like source spectrum, the estimated ideal sensitivity could achieve $1.9\times10^{-12}$\,erg~cm$^{-2}$~s$^{-1}$ for an exposure of 10$^4$\,s in the energy band of 0.5--4\,keV. Furthermore, after simulating the background caused by low-energy charged particles near the geomagnetic equator, we have determined that there is no need to install a magnetic deflector.}



\maketitle
\section{Introduction}\label{sec1}

The Chasing All Transients Constellation Hunters (CATCH) space mission is proposed to address the lack of follow-up observation capabilities in the time-domain astronomy era \cite{licatch}. It will study the dynamic universe via X-ray follow-up observations of various multi-wavelength and multi-messenger transients, such as electromagnetic counterparts of gravitational wave events, X-ray binaries, fast radio bursts, magnetars, and gamma-ray bursts. CATCH plans to consist of 126 X-ray micro-satellites that are controlled by an intelligent system. These satellites work together to perform various observations, including timing, spectroscopy, imaging, and polarization, for numerous transients simultaneously. The mission is composed of three types of satellites, each serving a different scientific purpose. Type-A satellites are used for immediate timing monitoring after target discovery. Based on the results obtained from type-A satellites, type-B satellites are deployed for more in-depth timing, imaging, and spectroscopic follow-up observations, and type-C satellites, on the other hand, are specifically designed for polarization measurements.

The current focus of the CATCH mission is on the development of its first Pathfinder, specifically designed for type-A satellites. For convenience, we will refer to the first pathfinder of CATCH as CATCH-1 in the following text. CATCH-1 has been designed to operate within the energy range of 0.5--4\,keV and is equipped with Micro Pore Optics (MPO) and a detector array comprising four Silicon Drift Detectors (SDD). Its primary objective is to conduct on-orbit verification of the MPO optics, SDD array detector system, deployable mast, and fast-pointing capability. Furthermore, it will demonstrate the sensitive observation capabilities of an X-ray telescope onboard a micro-satellite. CATCH-1 is currently being proposed for launch alongside the Space-based multi-band astronomical Variable Objects Monitor (SVOM) mission in 2024. If approved, both CATCH-1 and SVOM will share the same orbit, positioned at an altitude of 625\,km and an orbital inclination of $29^\circ$, which will facilitate efficient coordinated observations. 

Simulation studies play a crucial role in the development of CATCH-1, serving as an important step prior to its launch into orbit. These studies will help evaluate the performance capabilities of the satellite and verify if it can achieve its scientific objectives. They also aid in optimizing the shielding design of the detector, satellite configuration, and determining the necessity for a magnetic deflector. Additionally, simulation results can be used for ground calibration and provide valuable information for on-orbit observation plans and scientific data analysis. In this paper, we use the Geant4 Monte Carlo toolkit (version 4.10.06.p03), developed by the international Geant4 Collaboration, to carry out simulation studies of the scientific performance of the optical system, the on-orbit background, and telescope sensitivity. The selection of the Geant4 software is based on its capability to construct intricate geometric models and simulate particle-matter interactions \cite{1geant4,2geant4,3geant4}. Similar simulation approaches have been employed in various space satellites, including Swift \cite{swift}, XMM-Newton \cite{XMMNewton}, Insight-HXMT \cite{1HXMT,2HXMT}, and EP \cite{wxtEP,fxtEP}. These simulations have proven to be valuable in assessing the performance of these satellites and have contributed to their successful missions.

The paper is organized as follows: Section~\ref{sec2} introduces the configuration of CATCH-1, Section~\ref{sec3} describes the radiation environment in CATCH-1's orbit, Section~\ref{sec4} provides a comprehensive description of the simulation procedure, including the establishment of a mass model, the definition of physical processes, the generation of primary particles, and data processing, Section~\ref{sec5} presents the results of simulation, and Section \ref{sec6} concludes with a summary and discussion.

\section{Overall design}\label{sec2}

\begin{figure}[h]%
\centering
\includegraphics[width=0.87\textwidth]{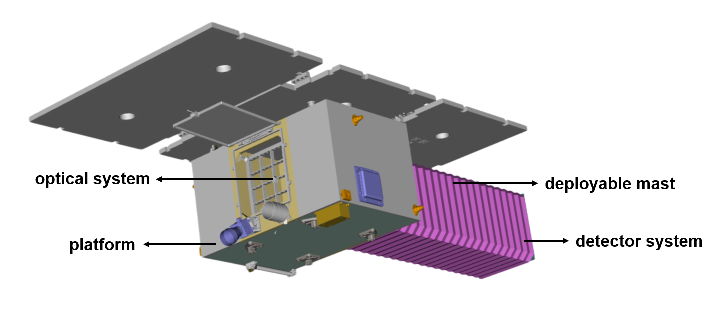}
\caption{\centering The CAD drawing of CATCH-1.}\label{fig1}
\end{figure}

Figure~\ref{fig1} shows the configuration of CATCH-1. It comprises a platform, a deployable mast, an optical system, and a detector system. The optical system is mounted in the front of the platform, and the detector system is mounted at the end of the deployable mast. The deployable mast is folded into the platform to minimize payload volume during launch. Once the CATCH-1 reaches its designated orbit, the deployable mast unfolds, precisely positioning the detectors at a focal distance of 1\,m from the focusing mirror. The total mass of CATCH-1 is $\sim$40\,kg.

\begin{figure}[h]%
\centering
\includegraphics[width=0.7\textwidth]{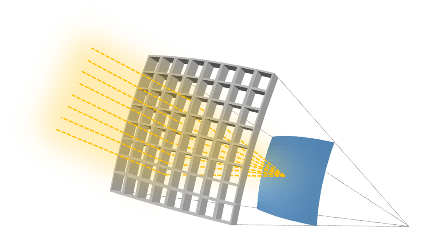}
\caption{\centering The structure of MPO optics and its focusing process.}\label{fig2}
\end{figure}

\begin{figure}[h]%
\centering
\includegraphics[width=0.95\textwidth]{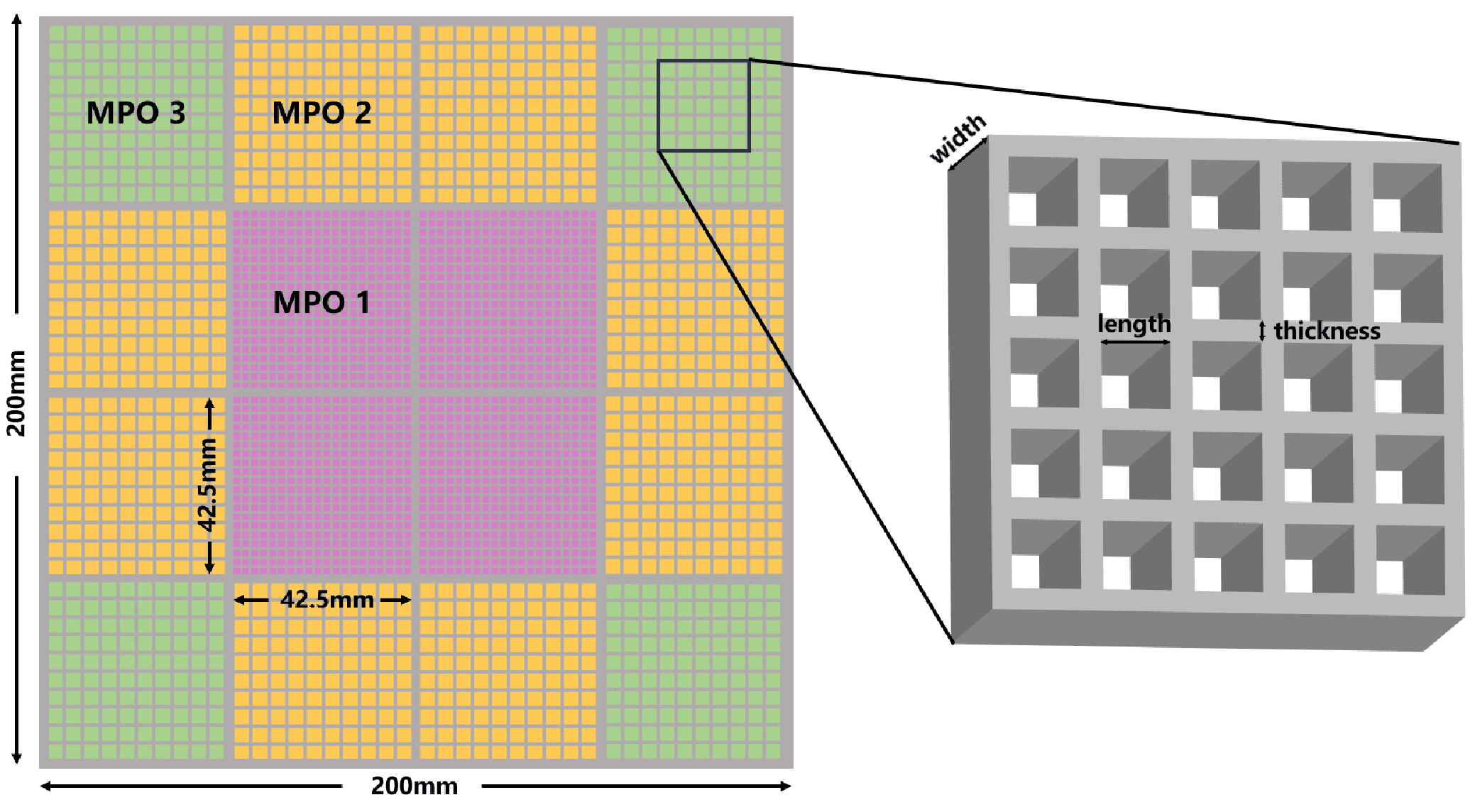}
\caption{(\textit{left}) The arrangement of the 4$\times$4 array of MPO mirrors in CATCH-1 (the size of micro pores is exaggerated for display). Three different colors represent three types of mirrors with different width-to-length ratios. We refer to the mirrors marked in purple as MPO\,1, the mirrors marked in yellow as MPO\,2, and the mirrors marked in green as MPO\,3. Their parameters are listed in Table \ref{tab1}. (\textit{right}) The enlarged view of the black border in the left panel.}\label{fig3}
\end{figure}

The optical system of CATCH-1 is MPO optics, also known as Lobster Eye X-ray Optics, which was initially proposed by Angel~\cite{angel}. This innovative design, inspired by the structure of lobster eyes, enables a larger field of view (FOV) compared to conventional Wolter-\uppercase\expandafter{\romannumeral1} X-ray optics. MPO optics is being used by several missions, such as EP~\cite{EPmirror}, SVOM~\cite{SVOMmirror}, and SMILE~\cite{SMILEmirror}. Figure \ref{fig2} illustrates the schematic of the structure of MPO optics and its focusing process. MPO optics is a spherical shell with millions of micro square pores that radially point towards a common center. X-rays are reflected off the sides of these pores and focused onto a spherical focal surface. The focal surface shares the same center as the optics and its curvature is half that of the optics. In the CATCH-1 mission, the optical system consists of a 4$\times$4 array of light-weight MPO mirrors (Figure \ref{fig3}), each measuring 42.5\,mm$\times$42.5\,mm. The total mass of the optical system is less than 0.75\,kg, and its overall size is 200\,mm$\times$200\,mm. Each mirror has a curvature of 2\,m and the focal length of the optical system is 1\,m. The FOV of the optical system is $0.4^\circ\times0.4^\circ$, making it the smallest FOV achieved by MPO optics to date. It is noteworthy that the MPO mirror array on CATCH-1 adopts the narrow-field-optimized lobster eye design in order to take the advantage of a larger effective area and better sensitivity~\cite{lobster}. This is achieved by adjusting the width-to-length ratio of the mirrors in the MPO mirror array. Specifically, the mirrors in the center of the array have a larger width-to-length ratio, while those located at the edges have a smaller ratio. The MPO mirror array of CATCH-1 consists of three types of mirrors with different width-to-length ratios, as shown in Figure \ref{fig3}. The parameters of each type are listed in Table \ref{tab1}. To enhance the performance of the optics, two layers of light-blocking films are coated on the mirror's surface. The first layer is a 50\,nm-thick organic film, which serves a dual function of shielding ultraviolet photons and providing structural support. The second layer is an 80\,nm-thick Aluminum film, primarily designed to block visible light and also contribute to the thermal control system. Additionally, a layer of Iridium is plated on the sides of the micro pores to improve the reflectivity of X-rays. The Point Spread Function (PSF) of the MPO optics is cross-shaped. We will provide a detailed explanation along with simulated images in Section \ref{sec5}.

\begin{table}[h]
\caption{The parameters of different types of MPO mirrors. }\label{tab1}%
\begin{tabular}{@{}ccccccc@{}}
\toprule
type &  number & width(mm) & length($\upmu$m) & thickness($\upmu$m) &  width-to-length ratio\\
\midrule
MPO\,1    &  4  &  1.25  &  20  &  5  &  62.5  \\
MPO\,2    &  8  &  1.25  &  40  &  8  &  31.3  \\
MPO\,3    &  4  &  1.1  &  40  &  8  &  27.5  \\
\botrule
\end{tabular}
\end{table}

\begin{figure}[h]%
\centering
\includegraphics[width=0.8\textwidth]{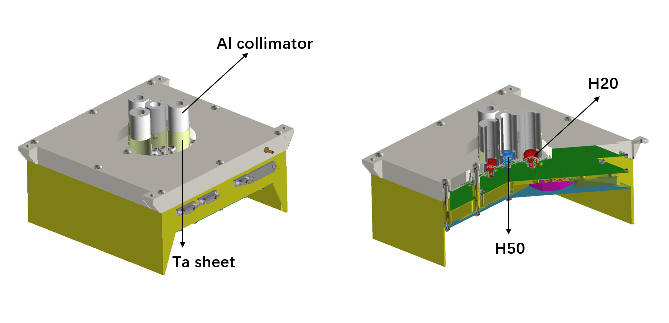}
\caption{The arrangement of the detector system (\textit{left}) and its cross section (\textit{right}). Each detector is placed inside a collimator with a height of 49.2\,mm to protect it from particles from the surroundings. The detector marked in blue is the H50 detector, and the other detectors marked in red are H20 detectors.}\label{fig4}
\end{figure}

The detector system of CATCH-1 is located at the end of the deployable mast. It consists of four SDD detectors manufactured by KETEK, arranged as shown in Figure \ref{fig4}. The selection of the SDD detector is based on its excellent time and energy resolution capabilities. The central detector in the array is the H50 detector, serving as the primary detector with a geometry area of 50\,mm$^2$. The three detectors surrounding it are H20 detectors, with two used for positioning and the remaining one used to measure background, each with a geometry area of 20\,mm$^2$. The total geometry area of the detector system is 110\,mm$^2$. The sensitive layers of these detectors are 450\,$\upmu$m-thick Silicon. The window material of these detectors is AP3.3 made by MOXTEK, which ensures maximum transmission of low-energy X-rays. Each detector is housed within an Aluminium collimator with a height of 49.2\,mm and a thickness of 5.0\,mm. This height allows for an effective reduction of the particle background from the surroundings while guaranteeing an unobstructed optical pathway within the aperture. On the exterior of the Aluminium collimator, there is a layer of cylindrical Tantalum sheet with a height of 19.2\,mm and a thickness of 0.5\,mm.

\section{Orbital environment}\label{sec3}

Accurate characterization of the space radiation environment in the satellite's orbit is essential for simulation studies, as it serves as a crucial input in simulating the satellite's on-orbit background. CATCH-1 is a low-orbit satellite with an orbital altitude of 625\,km and an orbital inclination of $29^\circ$ if launched alongside SVOM. In the subsequent simulations, we will utilize these orbital parameters. The radiation components that require consideration mainly include the Cosmic X-ray Background (CXB), albedo radiation, primary cosmic rays, secondary cosmic rays, and low-energy charged particles near the geomagnetic equator. In the following section, we will elaborate on these components and their corresponding spectral models.

\subsection{Cosmic X-ray background}\label{subsec2}

The CXB comes from the superimposed contribution of enormous extragalactic X-ray sources~\cite{CXBtheory}. The energy spectrum of CXB is generally defined as a broken power-law spectrum, as given by Equation \ref{eq1}~\cite{CXBequ},

\begin{equation}
F(E)=
\begin{cases}
0.54 \times E^{-1.4}, & E<0.02 \,\rm MeV,\\
0.0117 \times E^{-2.38}, & 0.02 \,{\rm MeV} \leq E<0.1 \,\rm MeV,\\
0.014 \times E^{-2.3}, & E \geq 0.1 \,\rm MeV.
\end{cases}\label{eq1}
\end{equation}
The flux is in units of counts~s$^{-1}$~cm$^{-2}$~sr$^{-1}$~MeV$^{-1}$.
Because of the decline of detection efficiency for high-energy photons and the exponential decrease in CXB flux with increasing energy, we only consider particles within the energy range of 0.1\,keV to 100\,MeV for sampling.

\subsection{Albedo radiation}\label{subsec3}

\begin{figure}[h]%
\centering
\includegraphics[scale=0.9]{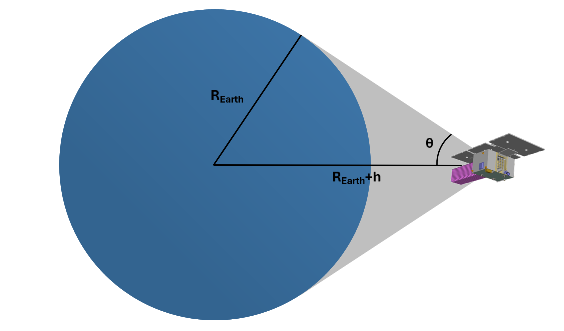}
\caption{\centering Albedo radiation enters from the shaded region. }\label{fig5}
\end{figure}

Albedo particles originate from the interaction of cosmic particles with the Earth's atmosphere, and the atmosphere reflection of cosmic particles. In this paper, we considered two components, albedo $\gamma$-ray and albedo neutrons. Albedo particles enter only from the direction of the Earth, as shown in the shaded region of Figure \ref{fig5}. The critical angle $\theta$ is calculated as $\theta = { \rm arcsin}(R_{\rm Earth}/(R_{\rm Earth}+h))$, where $R_{\rm Earth}$ = 6371\,km denotes the Earth's radius and $h$ = 625\,km represents the orbital altitude.

For albedo $\gamma$-ray, the energy spectrum of albedo $\gamma$-ray can be described by Equation \ref{eq2}~\cite{albedoequ},

\begin{equation}
F(E)=\displaystyle{\frac{0.0148}{{\left(\displaystyle{\frac{E}{33.7\,\rm keV}}\right)}^{-5}+{\left(\displaystyle{\frac{E}{33.7\,\rm keV}}\right)}^{1.72}}}.\label{eq2}
\end{equation}
The flux is in units of counts~s$^{-1}$~cm$^{-2}$~sr$^{-1}$~keV$^{-1}$.
The energy range for sampling is 10\,keV--10$^5$\,keV.

For albedo neutrons, the model prospected by Monte Carlo simulation~\cite{neutronfit} for the orbit of an altitude of 450\,km and inclination angle of $42^\circ$ in the solar minimum is used due to its extended energy coverage and the accurate validation~\cite{neutron}.
The energy range for sampling is 10\,keV--10$^5$\,keV.

\subsection{Primary cosmic rays}\label{subsec4}

Primary cosmic rays come from high-energy charged particles outside the solar system, including protons, positrons, and electrons. The energy spectrum of the Galactic primary cosmic rays in space can be described by a power-law spectrum, as presented by Equation \ref{eq3}~\cite{mizunocosmic},
\begin{equation}
{\rm Unmod}(E)=A \left[ \frac{R(E)}{\rm GV} \right] ^{-a}.\label{eq3}
\end{equation}
For protons, $A$ = 23.9\,counts~s$^{-1}$~m$^{-2}$~sr$^{-1}$~MeV$^{-1}$, $a$ = 2.83. For electrons, $A$ = 0.65\,counts~s$^{-1}$~m$^{-2}$~sr$^{-1}$~MeV$^{-1}$, $a$ = 3.30. $E$ is the kinetic energy of particle, and $R$ is the rigidity of particle. Positrons have the same energy spectrum as electrons, but the ratio of their fluxes $e^+/(e^++e^-)$ is 0.11~\cite{primaryeratio}.

In the solar system, primary cosmic rays undergo modulation due to solar activity, which can be described by Equation \ref{eq4}~\cite{solar},
\begin{equation}
{\rm Mod}(E)={\rm Unmod}(E+\left|Z\right|e \phi ) \times \frac{(E+Mc^2)^2-(Mc^2)^2}{(E+ Mc^2+ \left|Z\right|e \phi)^2-(Mc^2)^2},\label{eq4}
\end{equation}
where $M$ is the mass of particle, $c$ is the speed of light, and $Z$ is the charge of particle. $\phi$ is the solar activity modulation parameter, which changes periodically with the solar activity. In the simulation, we set $\phi$ to 900\,MV based on the on-orbit observation time of CATCH-1.

Primary cosmic rays are also influenced by the Earth's magnetic field. The geomagnetic correction factor $C$ is described by Equation \ref{eq5}~\cite{mizunocosmic},

\begin{equation}
C = \frac{1}{(1+ R/R_{\rm cut})^{-r}}.\label{eq5}
\end{equation}
For protons, $r = 12$; for positrons and electrons, $r = 6$. The geomagnetic cutoff rigidity $R_{\rm cut}$ is defined by Equation \ref{eq6}~\cite{geomagnetic},

\begin{equation}
R_{\rm cut}=14.9 \times \left(1+ \frac{h}{R_{\rm Earth}} \right) ^{-2.0} \times ({\rm cos}\theta_{\rm M})^{4} \,\rm GV,\label{eq6}
\end{equation}
where $h$ = 625\,km, $R_{\rm Earth}$ = 6371\,km, and $\theta_{\rm M}$ is the geomagnetic latitude. For CATCH-1, with an orbital inclination of $29^\circ$, the range of geomagnetic latitudes is 0 to 0.7. To provide a conservative estimate of the satellite's scientific capability and background level, we choose the parameter value that maximizes the flux. Therefore, we set $\theta_{\rm M}$ to 0.7 in the simulation. 

Thus, the energy spectrum of primary cosmic rays near the Earth can be described by Equation \ref{eq7}~\cite{mizunocosmic},

\begin{equation}
\begin{split}
F(E)=A \left[ \frac{R(E+\left|Z\right|e \phi)}{\rm GV} \right] ^{-a} &\times \frac{(E+Mc^2)^2-(Mc^2)^2}{(E+ Mc^2+ \left|Z\right|e \phi)^2-(Mc^2)^2}\\
&\times \frac{1}{(1+ R/R_{\rm cut})^{-2}}.\label{eq7}
\end{split}
\end{equation}
The flux is in units of counts~s$^{-1}$~m$^{-2}$~sr$^{-1}$~MeV$^{-1}$. 
The energy range for sampling is 100\,MeV--10$^6$\,MeV.

\subsection{Secondary cosmic rays}\label{subsec5}

Secondary cosmic rays are generated from the interaction of primary cosmic rays with the Earth's atmosphere. The energy spectrum of secondary cosmic rays can be described by a broken power-law spectrum, as given by Equation \ref{eq8}~\cite{mizunocosmic},

\begin{equation}
F(E)=
\begin{cases}
F_0  \left( \displaystyle{\frac{E}{100 \,\rm MeV}} \right) ^{-1}, & 10{ \,\rm MeV} \leq E<100{ \,\rm MeV},\\
F_0  \left( \displaystyle{\frac{E}{100 \,\rm MeV}} \right) ^{-a}, & 100{ \,\rm MeV} \leq E< E_{\rm bk},\\
F_0  \left( \displaystyle{\frac{E_{\rm bk}}{100 \,\rm MeV}} \right) ^{-a} \left( \displaystyle{\frac{E}{E_{\rm bk}}} \right) ^{-b}, &  E \geq E_{\rm bk}.
\end{cases}\label{eq8}
\end{equation}
The flux is in units of counts~s$^{-1}$~m$^{-2}$~sr$^{-1}$~MeV$^{-1}$. The values of parameters $F_0$, $a$, $E_{\rm bk}$, and $b$ depend on the geomagnetic latitude. We select the parameter values that maximize the flux. For protons, the values of $F_0$, $a$, $E_{\rm bk}$ and $b$ are 0.1\,counts~s$^{-1}$~m$^{-2}$~sr$^{-1}$~MeV$^{-1}$, 0.87, 600\,GeV, 2.53, respectively. For electrons, the values of $F_0$, $a$, $E_{\rm bk}$ and $b$ are 0.3\,counts~s$^{-1}$~m$^{-2}$~sr$^{-1}$~MeV$^{-1}$, 2.2, 3\,GeV, 4.0, respectively. Positrons have the same energy spectrum as electrons, while the ratio of their fluxes $e^+/e^-$ is 3.3~\cite{seconderatio}. The energy range for sampling is 10\,MeV--10$^5$\,MeV.

\subsection{Low-energy charged particles near the geomagnetic equator}\label{subsec6}

The flux of low-energy charged particles is significantly enhanced near the geomagnetic equator~\cite{lowenergytheory}. The energy spectrum of low-energy protons can be expressed by the kappa function, as given by Equation \ref{eq9}~\cite{1lowenergyequ,2lowenergyequ},

\begin{equation}
F(E)=A \left[ 1+ \frac{E}{k E_{0}} \right] ^{-k-1}.\label{eq9}
\end{equation}
The flux is in units of counts~s$^{-1}$~cm$^{-2}$~sr$^{-1}$~keV$^{-1}$, with $A$ = 328\,counts~s$^{-1}$~cm$^{-2}$~sr$^{-1}$~keV$^{-1}$, $E_0 = 73$\,keV, $k = 3.2$, $E_0 = 22$\,keV. The energy range for sampling is 10\,keV--10$^4$\,keV.

The energy spectrum of low energy electrons can be described by a Maxwell function at lower energies and a power function at higher energies, as given by Equation \ref{eq10}~\cite{3lowenergyequ},

\begin{equation}
F(E)=
\begin{cases}
A \displaystyle{\frac{E}{E_0}} {\rm exp}\left( \displaystyle{-\frac{E}{E_{0}}} \right), & E<1000{ \,\rm keV},\\
A E^{-\gamma}, & E \geq 1000{ \,\rm keV}.
\end{cases}\label{eq10}
\end{equation}
The flux is in units of counts~s$^{-1}$~cm$^{-2}$~sr$^{-1}$~keV$^{-1}$, with $A$ = 200\,counts~s$^{-1}$~cm$^{-2}$~sr$^{-1}$~keV$^{-1}$, $E_0 = 73$\,keV, $\gamma = 2.9 $. The energy range for sampling is 10\,keV--10$^4$\,keV.

\begin{figure}[h]%
\centering
\includegraphics[width=1.0\textwidth]{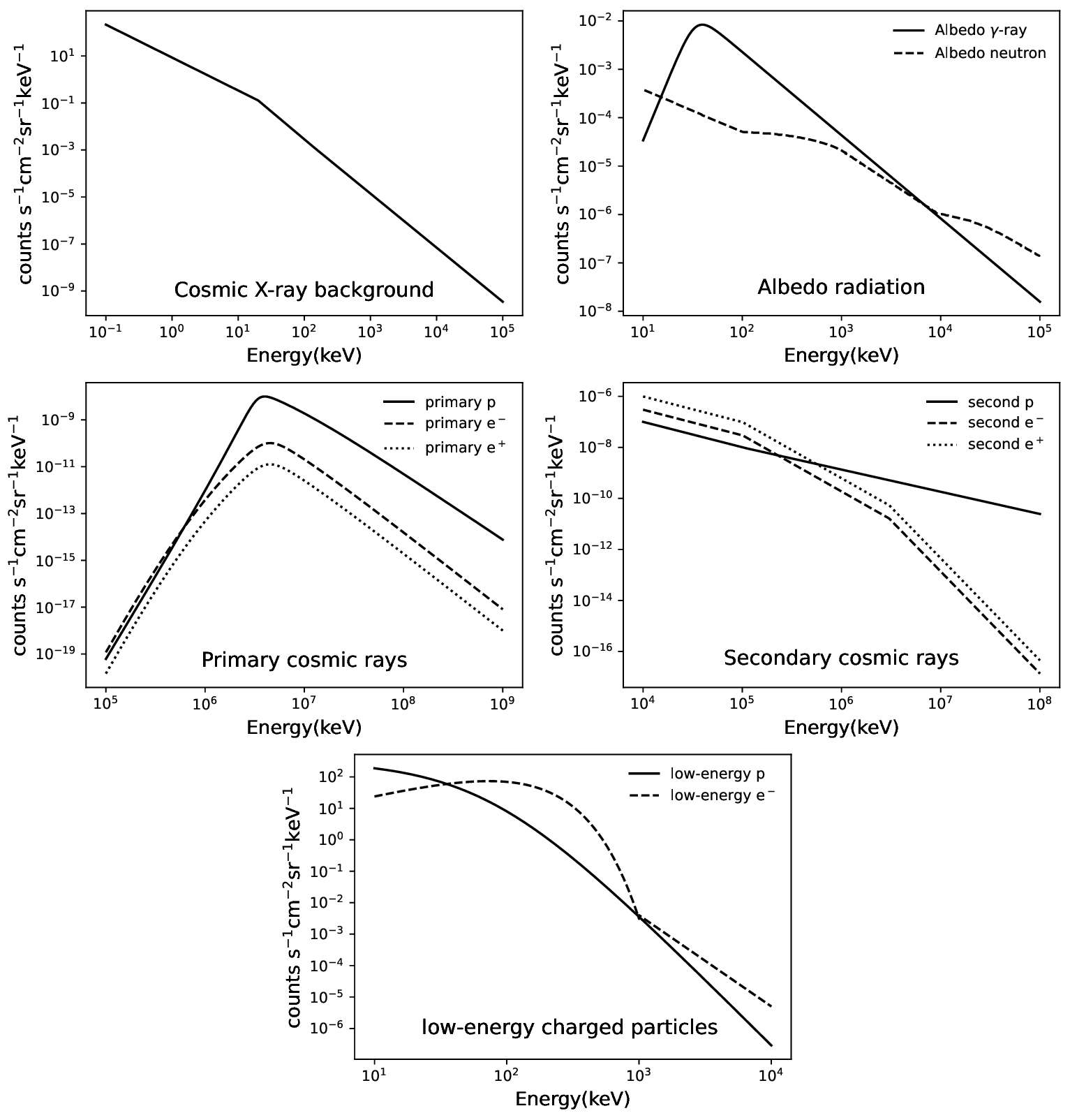}
\caption{Energy spectra of different types of particles in their respective energy ranges, used as input in Geant4 simulation of the background.}\label{fig6}
\end{figure}

The spectra of different types of particles within their respective energy ranges are plotted in Figure \ref{fig6}. It should be noted that low-energy charged particles only exist in the vicinity of the geomagnetic equator. Therefore, the background calculation focuses exclusively on the contributions from the CXB, albedo radiation, primary cosmic rays, and secondary cosmic rays. However, simulating the background from low-energy charged particles is indispensable as it provides a reference for determining the necessity for a magnetic deflector.

\section{Simulation}\label{sec4}

This section provides a comprehensive description of the simulation process, including the establishment of the CATCH-1 mass model, the definition of a physics list, the generation of primary particles, and data processing.

\subsection{Mass model}\label{subsec7}

\begin{figure}[ht]%
\centering
\includegraphics[width=1.0\textwidth]{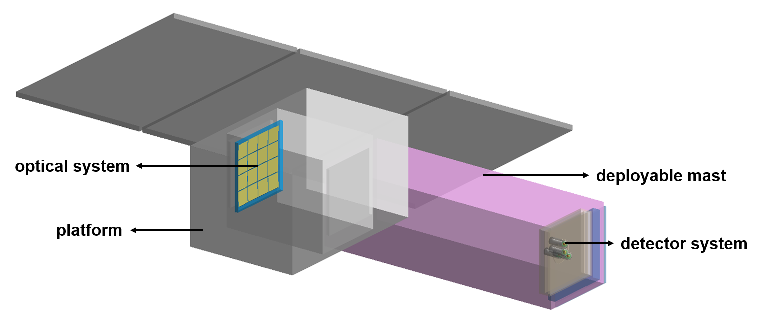}
\caption{\centering The mass model of CATCH-1 built in Geant4. }\label{fig7}
\end{figure}

In Geant4 simulations, establishing a satellite mass model with detailed definitions of geometry and material for each component is necessary. The Geant4 mass model for our simulation is shown in Figure \ref{fig7}. Section \ref{sec2} has provided a detailed description of the geometrical shapes and dimensions of the optical system and the detector system. Their geometrical configurations in Geant4 are consistent with the design specifications. In defining the materials for the optical system and detector system, we opted to represent them as pure substances, disregarding the impurities present in the actual materials, as these impurities have minimal impact on the results. For example, the material of the mirror, originally glass, has been defined as SiO$_2$, and the Aluminum alloy used for the collimator has been replaced with pure Aluminum in the simulation. For the platform and deployable mast, it is sufficient to ensure that their geometric shapes and total masses are consistent with the actual values. For example, the platform is simplified by assuming it is filled with Aluminum, with the Aluminum density determined by dividing the platform mass by the platform volume. Similarly, the deployable mast is simplified as a hollow rectangular box, disregarding the crease on its surface. These simplifications are justified by the fact that the intricate details of these structures have a negligible impact on the simulation results, and simplifying them can improve the simulation speed. 

When building the mass model of CATCH-1, in addition to what has been mentioned above, it is required to define the boundary surfaces where the grazing particles are scattered. We designate the border between the vacuum of the micro pores and the Iridium coating as the scattering surfaces. When tracking a particle passing through such surfaces, Geant4 will invoke the physics process of grazing angle scattering to simulate. Furthermore, we designate the sensitive layers of the four SDD detectors as sensitive detectors. This enables convenient tracking and recording of the energy deposition and position of particles within the detectors during simulation.

\subsection{Physics list}\label{subsec8}

The physics list describes the interaction between particles and the constituent materials of the satellite, encompassing all relevant physical processes. When the incident particles are neutrons, we implemented the Shielding Physics List provided by Geant4. When the incident particles are X-rays and charged particles, we created a Physics List from scratch and the electromagnetic physics constructor G4EmStandardPhysics\_option4 is chosen. The cut value is set to the default. The options of fluorescence (Fluo), particle induced X-ray emission (PIXE) and auger processes are all set to be active. Additionally, the scattering of X-rays and low-energy charged particles at grazing angles on the surfaces of micro pores is crucial for our simulation. However, Geant4 does not provide a corresponding physical model for this phenomenon. To address this, we adopt the physical model package developed by Qi et al.~\cite{qi} to describe the grazing angle scattering. This is achieved through the implementation of three specific classes~\cite{Xraytracing,zhao}: (1) G4DetectorConstruction defines the scattering surfaces for particles arriving at grazing angles. (2) G4XrayGrazingAngleScattering and G4ProtonGrazingAngleScattering describe the specific processes of the scattering of photons and protons, respectively. In this type of class, we need to provide the reflectivity of the boundary surfaces at different energies and incidence angles, which can be obtained from \url{https://cxro.lbl.gov//coatings}. The roughness of the Iridium film is set to 0.55 nm. (3) G4PhysicsList adds the G4XrayGrazingAngleScattering to the list of X-ray interactions and adds G4ProtonGrazingAngleScattering to the list of proton interactions. 

\subsection{Primary generation}\label{subsec9}

\begin{figure}[h]%
\centering
\includegraphics[width=1.0\textwidth]{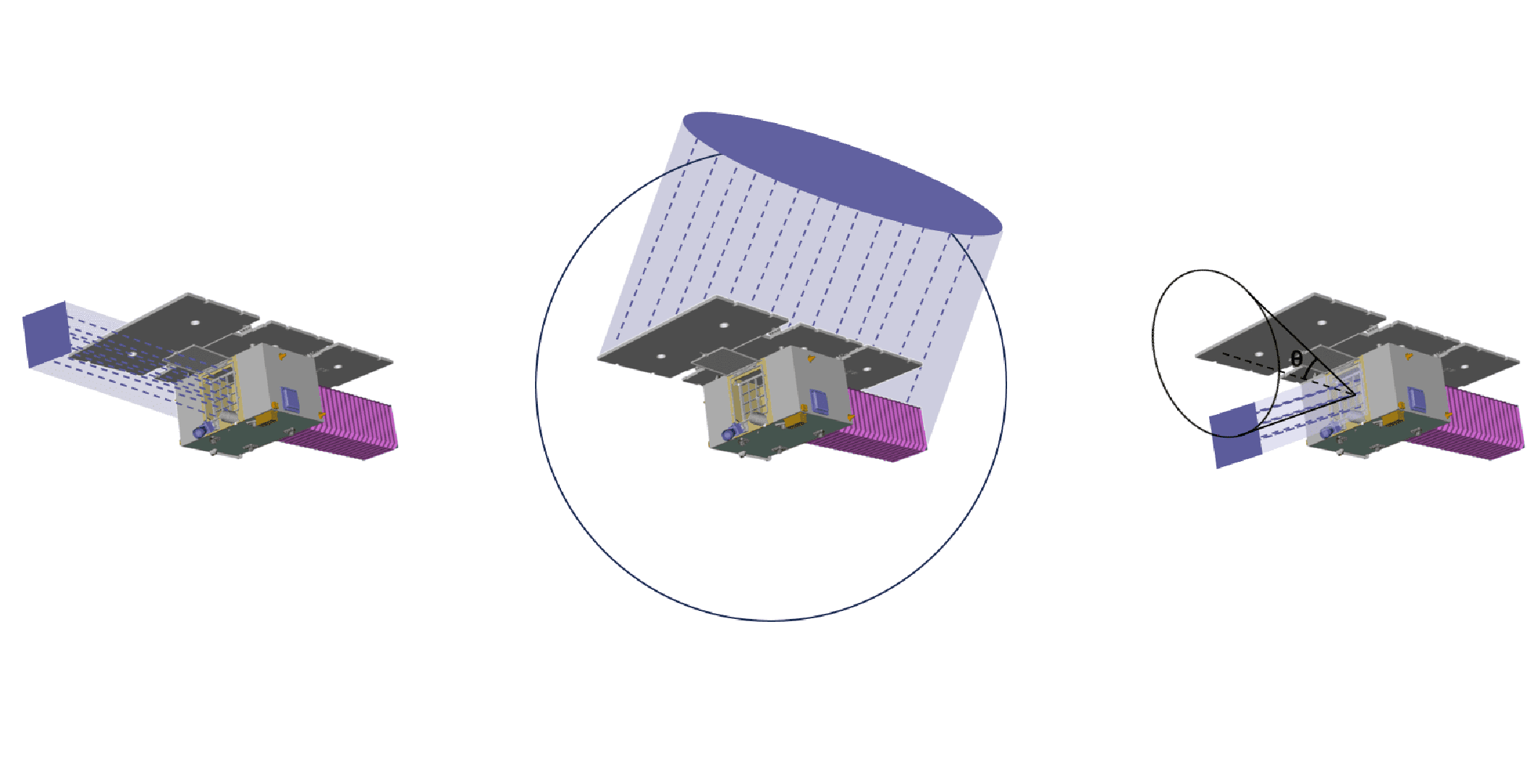}
\caption{Three modes of incidence: single-direction incidence (\textit{left}), isotropic incidence (\textit{middle}), and aperture incidence (\textit{right}). }\label{fig8}
\end{figure}

In simulations, we need to specify the particle type, energy, position, and direction of the primary source particles. In Section \ref{sec3}, we discussed the types of particles present in the orbital environment and their corresponding energy spectra. In this part, we will focus on the incident position and direction. Different simulation experiments require different modes of incidence sources, including single-direction incidence, isotropic incidence, and aperture incidence. Figure \ref{fig8} illustrates these modes.

We adopt single-direction incidence in simulating the properties of the MPO optics. The primary source particles are positioned directly in front of the mirror, taking the shape of a square plane that matches the overall size of the optical system and remains parallel to the mirror surface. All particles are incident in a parallel manner towards the mirror. The diagram of the single-direction incidence mode is illustrated in the left panel of Figure \ref{fig8}.

For simulating the background, as the radiation sources in the orbital environment are isotropically distributed, we employ isotropic incidence~\cite{fxtEP}. To sample the isotropic sources, firstly, a sphere that can encompass the entire satellite is chosen. To improve the simulation efficiency, the sphere should be as small as possible under the condition that it can envelop the satellite. We set the sphere's radius to 1\,m. Next, a point is randomly selected on this sphere based on the angle distribution of different particles. Taking this point as the center, a circle can be determined that is tangential to the sphere and of sufficient size to cover the cross-section of CATCH-1. We set the radius of this circle to 0.9\,m, which is half the diagonal length of the satellite's surrounding box. Finally, we uniformly select a point from this circle plane as the position of the primary particle. The incidence direction of this particle is from the center of the circle towards the center of the satellite. The diagram of the isotropic incidence is illustrated in the middle panel of Figure \ref{fig8}.

In the background simulation, we are concerned with the photon background inside the aperture. This is because photons inside the aperture are focused by the optical system, resulting in a significant contribution to the total background. In the simulation to determine the necessity for a magnetic deflector, we are interested in the background generated by charged particles inside the aperture. Due to the small region of the aperture, it is inefficient to adopt isotropic incidence and then pick out the particles from the aperture. Therefore, it is necessary to use aperture incidence in our simulation. The sampling method for aperture incidence is similar to that of isotropic incidence. However, there are two distinctions. Firstly, angle of incidence is no longer sampled from all directions on the spherical surface but is limited to within the aperture, as depicted by the solid black line in the right panel of Figure \ref{fig8}. The maximum value of incidence angle $\rm{\theta}$ is determined to be $6^\circ$ through simulation experiments. Secondly, the incidence source plane is no longer circular but a square with the size matching the outer envelope of the mirror. The incidence direction of this particle is from the center of the square plane towards the center of the mirror. The diagram of the aperture incidence is illustrated in the right panel of Figure \ref{fig8}.

\subsection{Data processing}\label{subsec10}

The incident particles interact with the satellite's material and lose energy in it. We record the energy deposited in detectors along with their corresponding spatial coordinates and export them for subsequent data analysis.

The first step is to broaden the deposited energy according to the energy resolution of the detector. 

The second step is normalization~\cite{1data,2data}. For each background component, we count the distribution of the deposited energy that has been broadened and generate an energy spectrum. The counts $M$ for each bin in the spectrum is given in counts~keV$^{-1}$. To convert the units to counts~keV$^{-1}$~s$^{-1}$, the spectrum needs to be normalized. The flux of the particle source, as described in Section \ref{sec3}, is defined as counts per unit time per unit area per unit solid angle per unit energy. By integrating the differential flux over the incident area, solid angle, and energy range in the simulation, the particle counts per unit time ($P\,\text{counts~s$^{-1}$}$) in the real space environment can be calculated, as given by Equation \ref{eq11},

\begin{equation}
P= \int dA \int d\Omega \int f(E)dE.\label{eq11}
\end{equation}
In order to obtain sufficient statistics in the simulation, the incident particle number $N$ is determined individually. Dividing $N$ by the particle counts $P$ gives the observation time $T$ in the simulation, as described in Equation \ref{eq12},

\begin{equation}
T= \frac{N}{P}.\label{eq12}
\end{equation}
By dividing the value $M$ of each bin in the spectrum by the observation time $T$, the unit can be normalized to counts~keV$^{-1}$~s$^{-1}$. 

\begin{figure}[h]%
\centering
\includegraphics[width=1.0\textwidth]{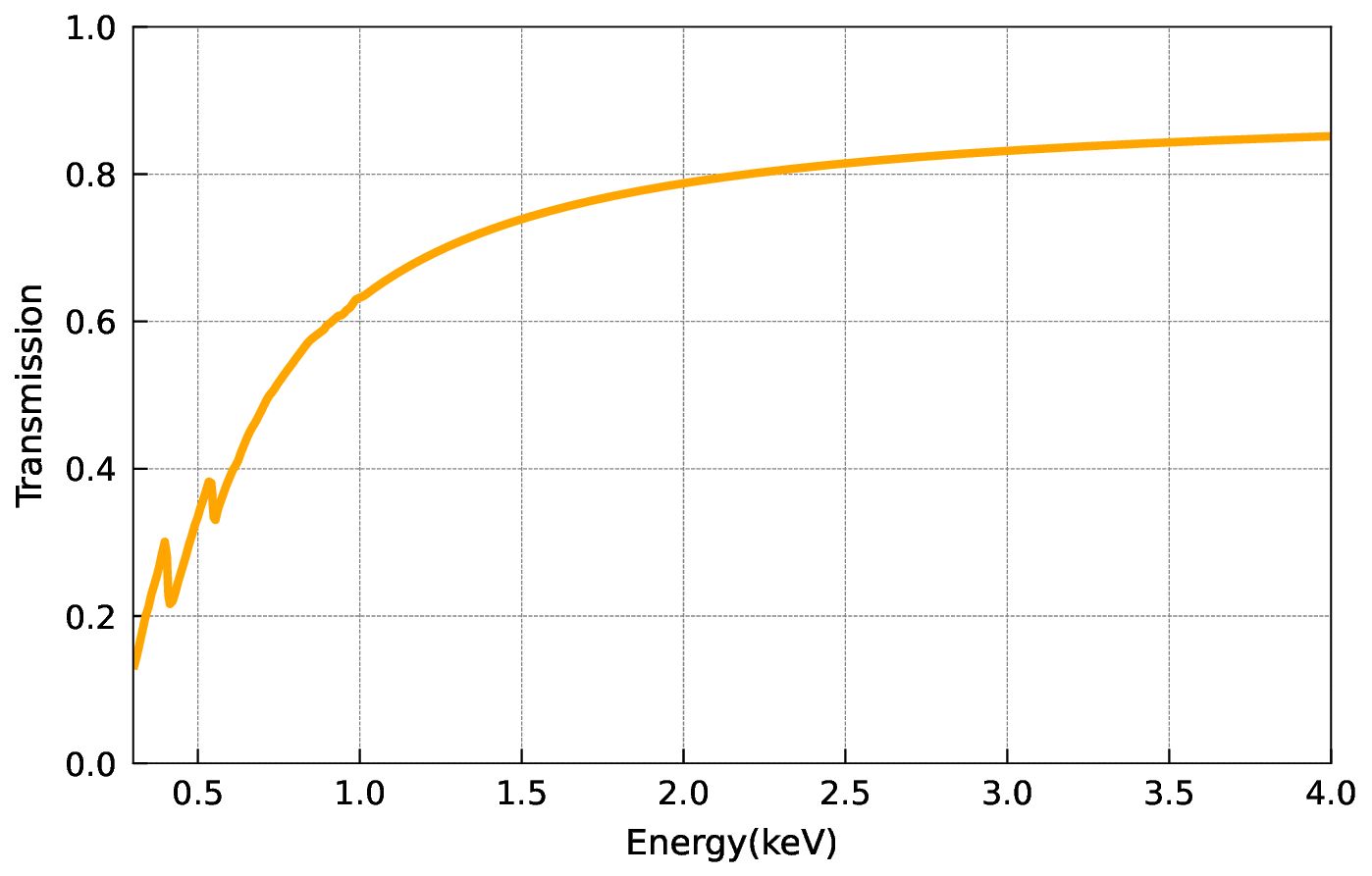}
\caption{\centering The X-ray transmission of the detector window made of AP3.3~\cite{AP3}. }\label{fig9}
\end{figure}

The detectors in CATCH-1 are equipped with windows made of the material AP3.3. However, due to the unavailability of detailed information regarding the material composition, the structure of the detector window was not included in the mass model. During data analysis, it is necessary to consider the impact of the detector window's transmission on the count rate. Fortunately, the X-ray transmission of AP3.3 is known~\cite{AP3}, as shown in the Figure \ref{fig9}. In order to obtain the final results, the counts of photons entering from the aperture must be multiplied by the transmission factor $R$.

Finally, the background count rate $C$ can be calculated by Equation \ref{eq13},
\begin{equation}
C=
\begin{cases}
\sum\limits_{i=1}^n R_i\displaystyle{\frac{M_{i}}{T}}, & \text{photons from the aperture} ,\\
\sum\limits_{i=1}^n \displaystyle{\frac{M_{i}}{T}}, &\text{others},
\end{cases}\label{eq13}
\end{equation}
where $n$ is the number of bins in the spectrum. The total background is the summation of the background count rate of each component.

\section{Results}\label{sec5}

In this section, we present the results obtained from simulations, including the scientific performance of the optical system, the on-orbit background, and telescope sensitivity.

\subsection{Mirror}\label{subsec11}

\begin{figure}[h]%
\centering
\includegraphics[width=1.0\textwidth]{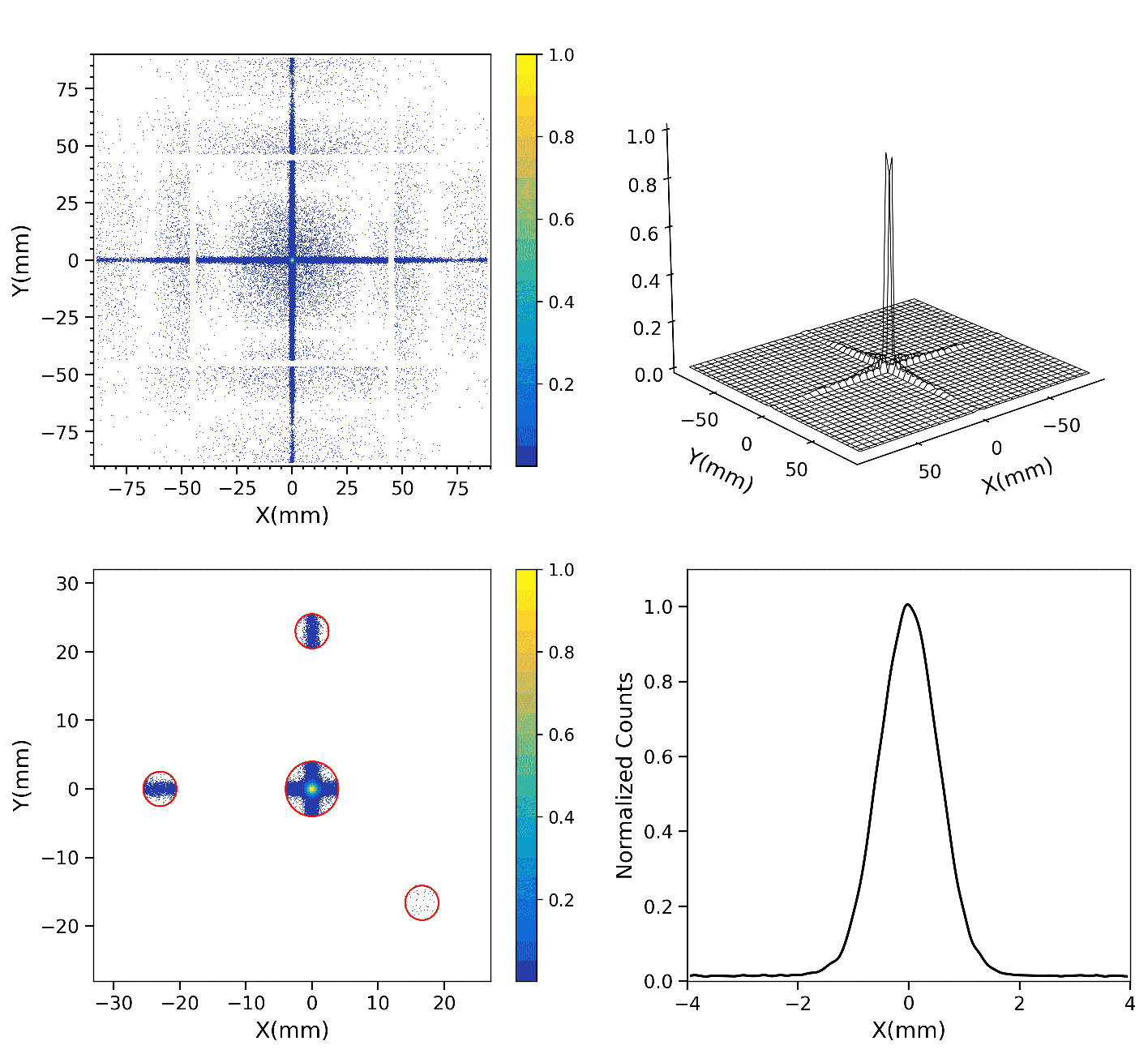}
\caption{The characteristic PSF of the optics of CATCH-1 got in the simulation. Each PSF image is obtained from 100,000 incident particles and the counts in these images have been normalized. (\textit{top left}) The PSF formed by the MPO mirror comprises a focused spot, horizontal and vertical cross-arms, and a diffuse patch. (\textit{top right}) The surface plot of PSF. (\textit{bottom left}) The PSF in the detector array of CATCH-1. (\textit{bottom right}) The one-dimensional PSF in the X direction.}\label{fig10}
\end{figure}

We get the characteristic Point Spread Function (PSF) of CATCH-1's optical system by simulation, as shown in Figure \ref{fig10}. During the simulation, we set the roughness of the reflecting surface to 0.55\,nm and assumed that the pointing deviation of the micro pores follows a Gaussian distribution with a mean value of 0 and $\sigma$ of 1.3\,arcmin. Under these conditions, the angular resolution is determined to be 4.5\,arcmin, consistent with experimental measurements. From the two images in the top panel of Figure \ref{fig10}, it can be observed that the cross feature formed by the MPO mirror comprises a focused spot, horizontal and vertical cross-arms, and a diffuse patch. If incident particles undergo two reflections off the sides of the micro pores, they will converge on the focal spot. Particles that undergo single or successive odd numbers of reflections will be deflected to the horizontal or vertical arms of the cross, while particles that pass straight through the micro pores or undergo multiple even numbers of reflections will fall onto the diffuse patch~\cite{lobster}. The gaps at $\pm$45\,mm are caused by the shadow of the supporting frame. The bottom left panel in Figure \ref{fig10} presents the PSF in the detector array of CATCH-1. It can be seen that the central detector precisely detected the focal spot, one positioning detector captured part of the horizontal cross-arm, another positioning detector captured part of the vertical cross-arm, and the background detector recorded partial diffuse patch. The positioning detectors can be utilized to improve the positioning accuracy of the telescope. The corresponding analysis is on-going and will be presented elsewhere. The bottom right panel in Figure \ref{fig10} displays the one-dimensional PSF in the X direction, which approximately follows a Gaussian distribution. 

\begin{figure}[h]%
\centering
\includegraphics[width=1.0\textwidth]{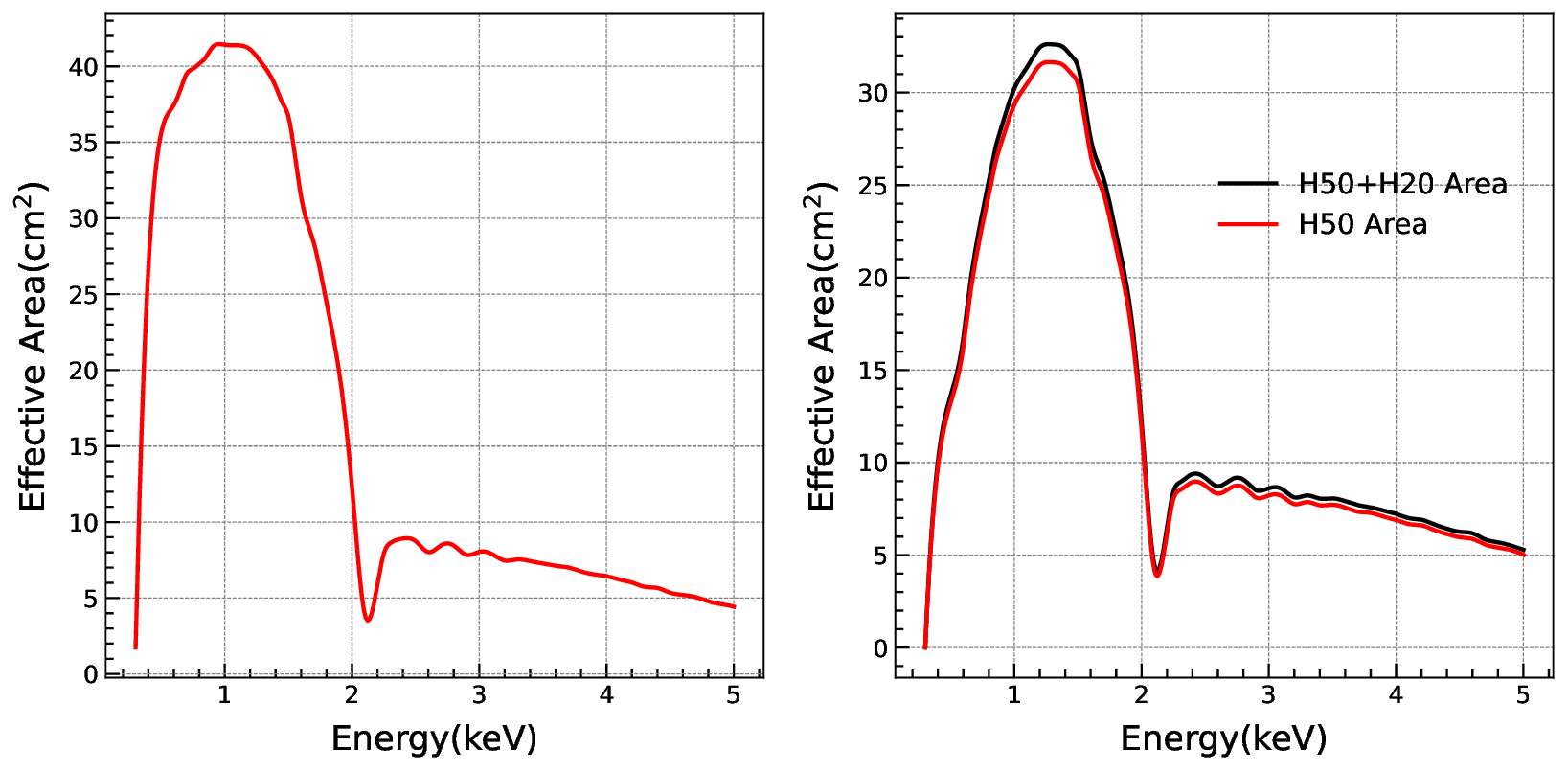}
\caption{(\textit{left}) For the individual optical system, the variations of the focal area (red) with energy. (\textit{right}) For the entire satellite, the variations of the effective area of the H50 detector (red) and that of all detectors (black) with energy. When calculating the effective areas of the entire satellite, the detection efficiency of the detector is taken into account.}\label{fig11}
\end{figure}

The effective area of CATCH-1 is also obtained through simulation, as shown in Figure \ref{fig11}. The left panel in Figure \ref{fig11} displays the variation of the focal area of the individual optical system with energy. For the individual optical system, the effective area of the focal spot is 41\,cm$^2$ at 1\,keV. As the detectors used by CATCH-1 do not have the capability of positional resolution, we calculate the effective area of the H50 detector and that of all detectors (H50+H20) separately, as shown in the right panel of Figure \ref{fig11}. It should be noted that the effective area of the H50 detector is not equivalent to the effective area of the focal spot. In addition to the focal spot, the central H50 detector also detects part of the cross-arms and diffuse patch. The effective area in the H50 detector is 29\,cm$^2$ at 1\,keV and 30\,cm$^2$ at 1.3\,keV. When calculating the effective areas of the detectors, we take into account the transmission efficiency of the detector window. The peak of the graph depicting the variation of the effective area with energy is shifted from 1\,keV to 1.3\,keV. 

\begin{figure}[ht]%
\centering
\includegraphics[width=0.75\textwidth]{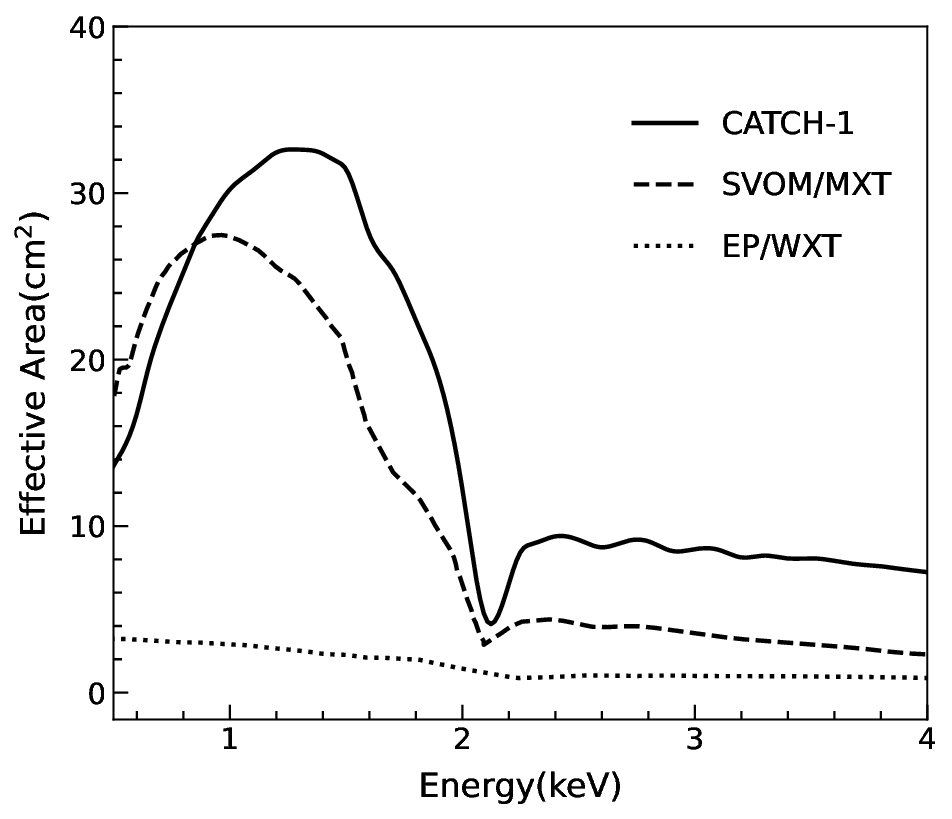}
\caption{The effective areas of CATCH-1, SVOM/MXT, and EP/WXT. The CATCH-1 area curve (solid line) is the effective area in the central H50 detector derived from the SDD detection efficiency. The SVOM/MXT area curve (dashed line) is the focal area derived from the pnCCD quantum efficiency, and an assumed 200 nm Al filter~\cite{SVOMmirror}. The EP/WXT area curve (dotted line) is the focal area for the individual optical system~\cite{wxtEP}.}\label{fig12}
\end{figure}

SVOM/MXT~\cite{SVOMmirror} and EP/WXT~\cite{wxtEP} are space missions that also use the lobster eye telescope. Figure \ref{fig12} shows the effective areas for CATCH-1, SVOM/MXT, and EP/WXT, respectively. It is evident that the effective area of CATCH-1 is comparable to that of SVOM/MXT, and above 0.9\,keV, CATCH-1 exhibits a larger effective area. The effective areas of CATCH-1 and SVOM/MXT are an order of magnitude larger than that of EP/WXT. This is because CATCH-1 and SVOM/MXT use the narrow-field-optimized lobster eye design, featuring a varying width-to-length ratio in the MPO mirrors array to maximize the on-axis effective area. In contrast, EP/WXT adopts a wide-field lobster eye design, maintaining a constant width-to-length ratio in the MPO mirror array to achieve a larger FOV but at the expense of a smaller effective area.

\subsection{Background}\label{subsec12}

\begin{figure}[h]%
\centering
\includegraphics[width=1.0\textwidth]{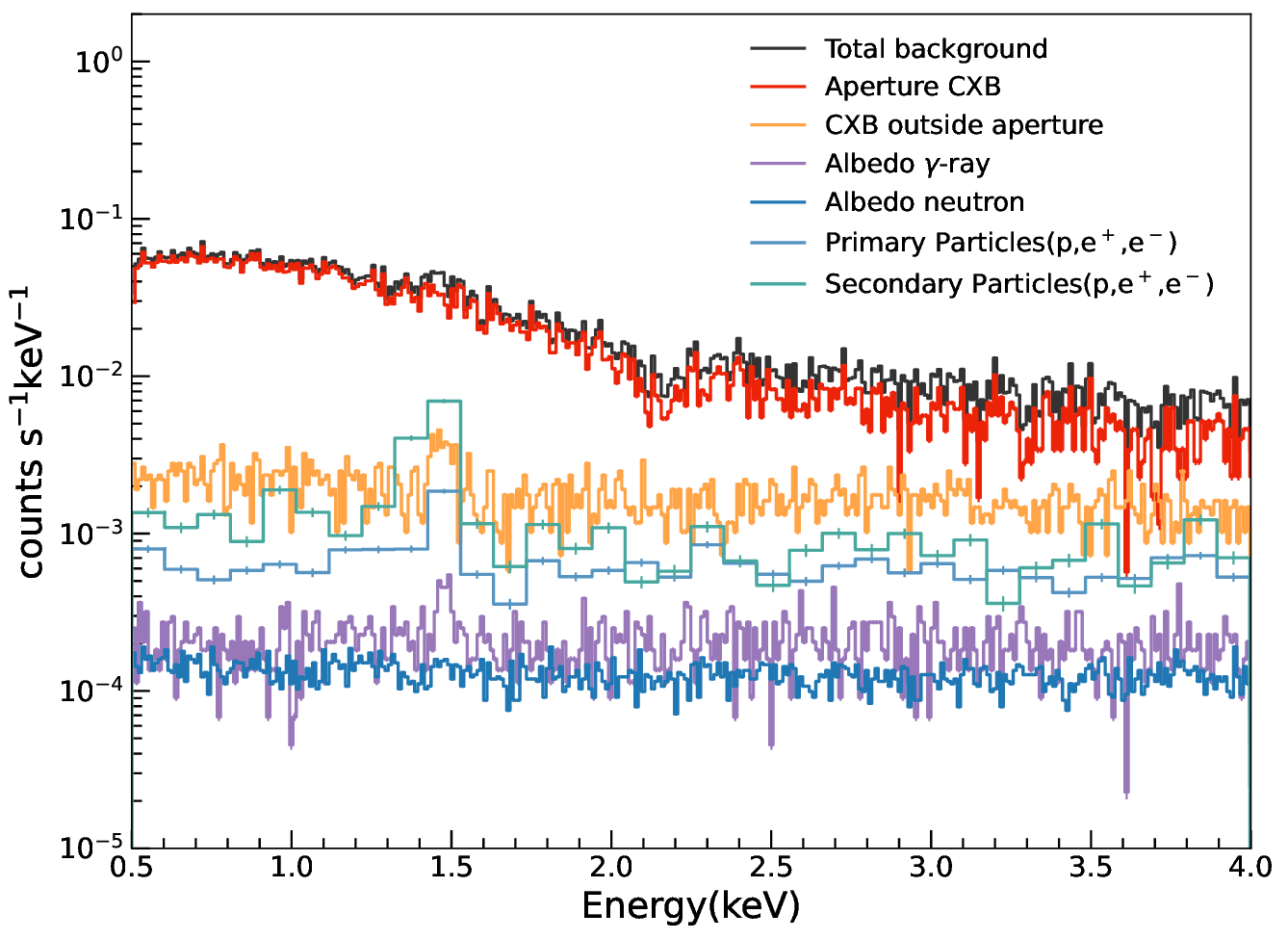}
\caption{The background energy spectra caused by various components, including CXB, albedo $\gamma$-ray, albedo neutrons, primary particles, and secondary particles. The CXB is divided into two parts: inside and outside the aperture. Primary particles refer to the primary cosmic rays, including protons, positrons, and electrons. Secondary particles refer to the secondary cosmic rays, including protons, positrons, and electrons. The total background spectrum, obtained by summing all these components, is represented by the black line.}\label{fig13}
\end{figure}

The background spectra caused by various components in the detectors on CATCH-1 are presented in Figure \ref{fig13}. In this figure, CXB is divided into two components, representing the contributions from inside the aperture and from outside the aperture, respectively. It is evident that the background of CATCH-1 is primarily contributed by the CXB inside the aperture. The shape of the spectrum of the CXB inside the aperture is correlated with the shape of the effective area of the mirror. The background spectra of the CXB outside the aperture, albedo $\gamma$-ray, albedo neutrons, primary particles, and secondary particles are relatively uniform between 0.5--4\,keV. The line between 1.4--1.5\,keV on their spectrum is associated with the K$_\alpha$ transition of Aluminum, which is caused by the reaction of particles with the collimator surrounding the detectors. Table \ref{tab2} is a detailed listing of the background from various components inside and outside the aperture. In the energy range of 0.5--4\,keV, the total background of the four SDD detectors in CATCH-1 is $8.13\times10^{-2}$\,counts~s$^{-1}$, with a statistical error of $1 \%$. The contribution from CXB inside the aperture accounts for 83.4\%. The total area of the detectors is 1.1\,cm$^2$, thus the background per unit area is $7.39\times10^{-2}$\,counts~s$^{-1}$~cm$^{-2}$.

\begin{table}[h]
\caption{The background in the four SDD detectors from various components inside and outside the aperture. The energy range is 0.5--4\,keV. The unit is counts~s$^{-1}$. }\label{tab2}%
\begin{tabular}{@{}ccccccc@{}}
\toprule
Particle &  Inside the aperture & Outside the aperture \\
\midrule
CXB    &  $(6.78\pm0.07)\times10^{-2}$  &  $(6.12\pm0.09)\times10^{-3}$    \\
Albedo $\gamma$-ray    &  -  &  $(7.03\pm0.12)\times10^{-4}$    \\
Albedo neutron    &  -  &  $(4.37\pm0.04)\times10^{-4}$    \\
Primary p    &   $(5.06\pm1.46)\times10^{-7}$  &   $(2.08\pm0.07)\times10^{-3}$  \\
Primary e$^+$    &  $(5.95\pm2.25)\times10^{-9}$  &    $(1.29\pm0.06)\times10^{-5}$   \\
Primary e$^-$    &  $(6.19\pm2.06)\times10^{-8}$  &    $(1.03\pm0.05)\times10^{-4}$  \\
Secondary p    &  $(2.10\pm0.48)\times10^{-7}$  &    $(5.41\pm0.18)\times10^{-4}$  \\
Secondary e$^+$    &  $(1.08\pm0.25)\times10^{-7}$  &   $(2.70\pm0.15)\times10^{-3}$   \\
Secondary e$^-$    &   $(6.46\pm1.72)\times10^{-7}$ &   $(8.17\pm0.40)\times10^{-4}$   \\
\botrule
\end{tabular}
\end{table}

We compare the background levels between CATCH-1 and SVOM/MXT, with SVOM/MXT serving as a reference due to its utilization of a narrow-field-optimized lobster eye telescope with a FOV of $1.06^\circ\times1.06^\circ$ and a 450\,$\upmu$m-thick Silicon sensitive layer in its detector, which is similar to the configuration of CATCH-1. The specific data are listed in Table \ref{tab3}~\cite{SVOMbackground}. It is evident that both the CXB background inside the aperture and the total background outside the aperture of CATCH-1 are at the same level as SVOM/MXT. This validates the reasonableness of the background simulation results.

\begin{table}[h]
\caption{Comparison of the background of CATCH-1 with that of SVOM/MXT. The unit is counts~s$^{-1}$~cm$^{-2}$~keV$^{-1}$. }\label{tab3}%
\begin{tabular}{@{}ccccccc@{}}
\toprule
       &  CXB background inside the aperture & Total background outside the aperture \\
\midrule
CATCH-1    &  $1.76\times10^{-2}$  &  $3.51\times10^{-3}$    \\
SVOM/MXT    &  $1.80\times10^{-2}$  &   $9.50\times10^{-4}$   \\
\botrule
\end{tabular}
\end{table}

We conduct a simple analysis in the following to illustrate whether it is necessary to install a magnetic deflector on CATCH-1. On the one hand, the total background of CATCH-1 is $8.13\times10^{-2}$\,counts~s$^{-1}$, while the contribution from primary and secondary particles inside the aperture is $1.54\times10^{-6}$\,counts~s$^{-1}$. Thus, the background from the charged particles inside the aperture can be neglected. On the other hand, for low-energy charged particles at the geomagnetic equator, the background inside the aperture is $1.4\times10^{-1}$\,counts~s$^{-1}$, whereas the background outside the aperture is 10.6\,counts~s$^{-1}$. Due to the background from low-energy charged particles inside the aperture being two orders of magnitude lower than that outside the aperture, the installation of a magnetic deflector behind the mirror would have little impact on the background. Furthermore, low-energy charged particles only exist in the vicinity of the geomagnetic equator. Therefore, the current conclusion is that the installation of a magnetic deflector is unnecessary.

\subsection{Sensitivity}\label{subsec13}

Sensitivity represents the minimum flux that a detector can detect. It is an important indicator for evaluating the observing capability of a telescope. The sensitivity of CATCH-1 can be estimated based on the simulated background. In this study, the ideal scenario where there is no uncertainty in the background is adopted~\cite{sensitivity}. In this case, the minimum counts $M$ that a source needs to produce in a detector in order to be detected above a 5$\sigma$ significance level in one observation can be calculated by a given background $B$, as described in Equation \ref{eq14},
\begin{equation}
M= a+b\sqrt{B}\label{eq14}
\end{equation}
where $a=11.090$, $b=7.415$. The background count $B$ is the background in the central detector H50, as the focal spot is in it. The total background in the central detector is $4.95\times10^{-2}$\,counts~s$^{-1}$ in the energy range of 0.5--4\,keV. The sensitivity is computed by assuming the observation of a Crab-like source. The Crab-like source spectrum, which is an absorbed power-law spectrum with an index of 2.05 and a column density $N_{\rm H} = 2\times10^{21}$\,cm$^{-2}$~\cite{wxtEP}, is obtained from XSPEC~\cite{xspec}. With an exposure time of 10$^3$\,s and 10$^4$\,s, CATCH-1 is able to achieve a sensitivity of $7.1\times10^{-12}$\,erg~cm$^{-2}$~s$^{-1}$ and $1.9\times10^{-12}$\,erg~cm$^{-2}$~s$^{-1}$ in the energy band of 0.5--4\,keV, respectively. This result is consistent with our expectations and fulfills the scientific requirements.

\begin{figure}[h]%
\centering
\includegraphics[width=1.0\textwidth]{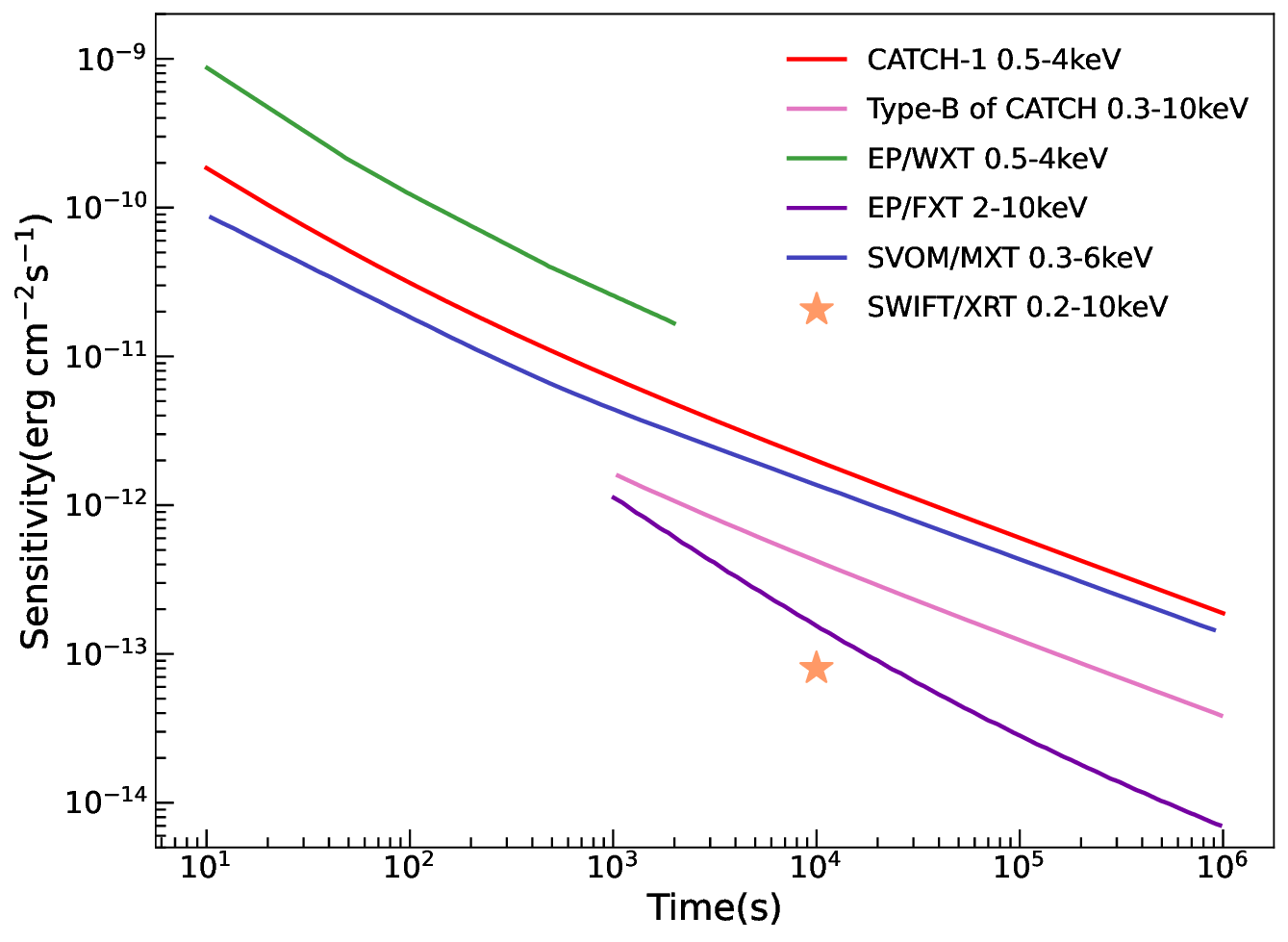}
\caption{The sensitivity of CATCH-1. The sensitivities of type-B satellites of CATCH, EP/WXT, EP/FXT, SVOM/MXT, and SWIFT/XRT are plotted on the same figure for comparison. }\label{fig14}
\end{figure}

The sensitivity of CATCH-1 is presented in Figure \ref{fig14}. For comparison, the sensitivities of the type-B satellites of CATCH~\cite{xiaocatch}, EP/WXT~\cite{wxtEP}, EP/FXT~\cite{fxtEP}, SVOM/MXT~\cite{SVOMmirror} and SWIFT/XRT~\cite{senSwiftXRT} are also plotted on the same figure. CATCH-1 has better sensitivity when compared to EP/WXT, which also uses the lobster eye telescope. Table \ref{tab4} lists the effective areas and sensitivities of CATCH-1 and EP/WXT, and their corresponding ratios. It is evident that the square of the sensitivity ratio is inversely proportional to the effective area ratio, which is consistent with theory~\cite{sensitivitytheory}. This supports the reasonableness of the estimated sensitivity for CATCH-1 in this study. When comparing SVOM/MXT with CATCH-1, SVOM/MXT exhibits better sensitivity due to its utilization of pnCCDs with 256$\times$256 pixels. However, the SDDs used by CATCH-1 outperform pnCCDs in terms of readout capability, which effectively prevents saturation from intense sources and offers a wider range of flux intensity. Therefore, if CATCH-1 and SVOM are on the same orbit, CATCH-1 can effectively carry out coordinated and relay observations with SVOM/MXT, thereby maximizing the strengths of each instrument. EP/FXT and SWIFT/XRT, both employing Wolter-\uppercase\expandafter{\romannumeral1} X-ray optics, possess stronger photon collection capabilities, resulting in better sensitivity than CATCH-1. In comparison with the Wolter-\uppercase\expandafter{\romannumeral1} optics, the MPO optics used by CATCH-1 is the light-weight design that can provide a larger effective area per kilogram of optics. This makes it more suitable for small satellite platforms. In the constellation layout of CATCH, there are also a few satellites using lightweight Wolter-\uppercase\expandafter{\romannumeral1} X-ray optic for in-depth timing, imaging, and spectroscopic observations, i.e. type-B satellite. Compared with CATCH-1 (for type-A satellite), type-B satellite has better sensitivity, reaching a level similar to EP/FXT and SWIFT/XRT (Figure~\ref{fig14}).

\begin{table}[h]
\caption{The effective areas and sensitivities of CATCH-1 and EP/WXT, and their corresponding ratios. The effective area is at 1\,keV. The sensitivity is at an exposure time of 10$^3$\,s.}\label{tab4}%
\begin{tabular}{@{}ccccccc@{}}
\toprule
       &  CATCH-1 & EP/WXT  & CATCH-1 : EP/WXT \\
\midrule
Effective Area\,(cm$^2$)    &  29.29  &  2.50  &  11.68 : 1    \\
Sentivity\,(erg~cm$^{-2}$~s$^{-1}$)    &  $7.14\times10^{-12}$  &  $2.40\times10^{-11}$   & 1 : 3.36 \\
\botrule
\end{tabular}
\end{table}

\section{Conclusion}\label{sec6}

CATCH is an intelligent constellation of 126 micro-satellites proposed for follow-up observations of substantial transients. In this work, we performed simulation studies of the first Pathfinder of CATCH using the Monte Carlo software Geant4. In Geant4, we established a mass model of the satellite, defined a physics list to describe the interaction between particles and the constituent materials of the satellite, and specified the particle type, energy, position, and direction of the primary source particles. The energy deposited in detectors and their corresponding spatial coordinates were recorded during the simulation for data analysis.

For the optics system, we adopted the physical model package developed by Qi et al~\cite{qi}. to describe the scattering of X-rays at grazing angles on the surfaces of micro pores. Through simulation, we obtained the characteristic PSF of the optics. The cross feature formed by MPO optics comprises a focused spot, horizontal and vertical cross-arms, and a diffuse patch. Furthermore, we calculated the effective area of the optics. For the individual optical system, the effective area of the focal spot is 41\,cm$^2$ at 1\,keV, while for the entire satellite, taking into account the transmission efficiency of the detector window, the effective area in the central detector is 29\,cm$^2$ at 1\,keV. 

In the simulation of the background, we considered various radiation components in CATCH-1's orbit, including CXB, albedo radiation, primary cosmic rays, and secondary cosmic rays. The total background of CATCH-1 is $8.13\times10^{-2}$\,counts~s$^{-1}$, and the background per unit area is $7.39\times10^{-2}$\,counts~s$^{-1}$~cm$^{-2}$. The primary background contribution comes from CXB inside the aperture, accounting for 83.4\%. Additionally, we simulated the background from low-energy charged particles near the geomagnetic equator and concluded that the magnetic deflector is not necessary.

Based on the background in the central detector H50, we estimated the sensitivity of CATCH-1. By assuming a Crab-like spectrum, with an exposure time of 10$^3$\,s, CATCH-1 can achieve a sensitivity of $7.1\times10^{-12}$\,erg~cm$^{-2}$~s$^{-1}$ in the energy band of 0.5-4\,keV. This result is consistent with our expectations and satisfies scientific requirements. 

In conclusion, our simulation studies have provided insights into the performance capabilities of the first CATCH-1. These simulation results will serve as input for ground calibration, formulation of on-orbit observation plans, and analysis of scientific data.

\section*{Acknowledgments}

We would like to express our gratitude to all colleagues in the CATCH team for their contributions throughout this work. We are also grateful for the support provided by Tencent. Furthermore, we would like to extend our thanks to Stéphane Schanne and Bertrand Cordier for their insightful suggestions regarding the arrangement of our detectors. This work is supported by the National Natural Science Foundation of China (NSFC) under the Grant Nos. 12122306, 12003037 and 12173056, the Strategic Priority Research Program of the Chinese Academy of Sciences XDA15016400, the CAS Pioneer Hundred Talent Program Y8291130K2. We also acknowledge the Scientific and technological innovation project of IHEP E15456U2.

\section*{Author Contributions}

Yiming Huang and Juan Zhang wrote the main manuscript. Lian Tao and Jingyu Xiao assisted with the critical revision of the article. All authors reviewed the manuscript and contributed to the development of the simulation studies for CATCH-1.

\section*{Funding}

We acknowledge funding support from the National Natural Science Foundation of China (NSFC) under the Grant Nos. 12122306 and 12003037, the Strategic Priority Research Program of the Chinese Academy of Sciences XDA15016400, the CAS Pioneer Hundred Talent Program Y8291130K2, and the Scientific and technological innovation project of IHEP E15456U2.

\section*{Conflict of interest}

The authors declare that they have no known competing financial interests or personal relationships that could have appeared to influence the work reported in this paper.

\bibliography{ref}

\end{document}